\documentclass[a4paper,
               biblatex,     
               keeplastbox,   
               ]{jacow}

\usepackage{pdfpages,multirow,ragged2e} %

\makeatletter%
	\ifboolexpr{bool{xetex}}
	 {\renewcommand{\Gin@extensions}{.pdf,%
	                    .png,.jpg,.bmp,.pict,.tif,.psd,.mac,.sga,.tga,.gif,%
	                    .eps,.ps,%
	                    }}{}
\makeatother

\ifboolexpr{bool{xetex} or bool{luatex}} 
 {}     
 {\usepackage[utf8]{inputenc}} 

\usepackage[USenglish]{babel}
\usepackage{pdflscape}  

\begin{document}

\title{Review of Technologies for Ion Therapy Accelerators}

\author{H. X. Q. Norman\thanks{hannah.norman-3@postgrad.manchester.ac.uk}\textsuperscript{1,2}, A. F. Steinberg\thanks{adam.steinberg@postgrad.manchester.ac.uk}\textsuperscript{{1,2}}, R. B. Appleby\textsuperscript{1}, University of Manchester, Manchester, UK \\
        H. L. Owen\textsuperscript{1}, STFC Daresbury, Warrington, UK\\
		E. Benedetto, M. Sapinski, SEEIIST Association, Geneva, Switzerland \\
		S. L. Sheehy, University of Melbourne, Victoria, Australia\\
		\textsuperscript{1}{also at Cockcroft Institute, Warrington, UK},
		\textsuperscript{2}{also at University of Melbourne, Victoria, Australia}\\
	}
	
\maketitle

\begin{abstract}
   Cancer therapy using protons and heavier ions such as carbon has demonstrated advantages over other radiotherapy treatments. To bring about the next generation of clinical facilities, the requirements are likely to reduce the footprint, obtain beam intensities above \num{1E10} particles per spill, and achieve faster extraction for more rapid, flexible treatment. This review follows the technical development of ion therapy, discussing how machine parameters have evolved, as well as trends emerging in technologies for novel treatments such as FLASH. To conclude, the future prospects of ion therapy accelerators are evaluated.
\end{abstract}

\section{Introduction} In recent years, using hadrons for radiotherapy has become more widely recognised for significant benefits with the sparing of healthy tissue \cite{Newhauser2015}. Carbon ions can be more beneficial than light ions when treating radio-resistant and deep-seated tumours, however the adoption of carbon ion therapy is hampered by the footprint of the accelerator and gantry, which are presently much larger than conventional radiotherapy devices \cite{Linz2012}. This has been the focus for improvement over the past 20+ years \cite{Gerbershagen2016}.

\section{Current and Developing Beam Delivery Systems}
\begin{figure}[!htb]
   \centering
   \includegraphics*[width=1\columnwidth]{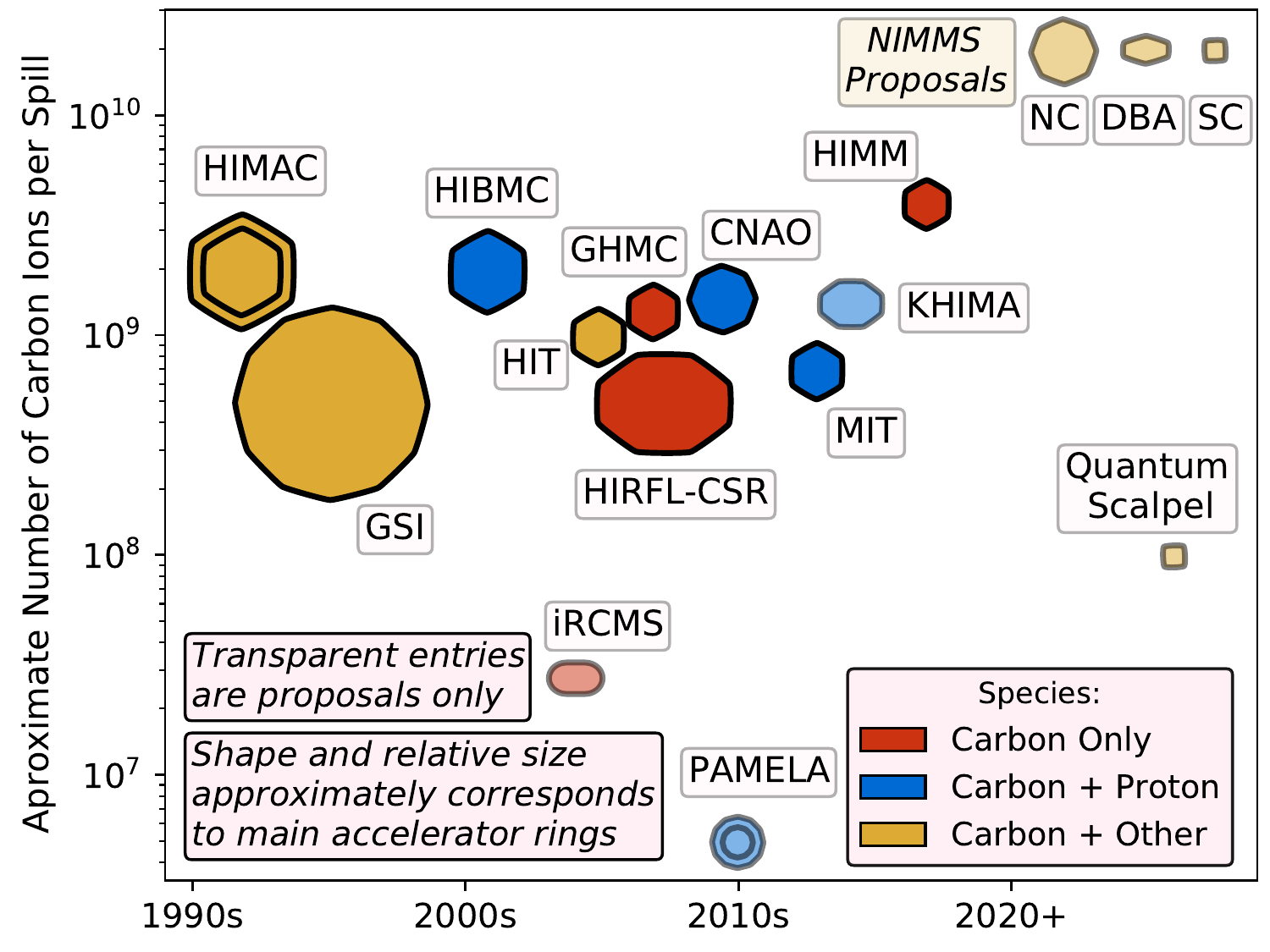}
   \caption{Progress in the design of ion therapy accelerators over time. Accelerators with relatively few particles per spill rely on technologies that allow for a higher repetition (cycle) rate to produce the same overall current, such as the use of static magnetic fields for PAMELA.}
   \label{fig:progress_vector}

\end{figure}

There have been many accelerators built for hadron therapy. A diagram of the progress of representative machines is shown in Fig.~\ref{fig:progress_vector}. The general trend is that the size of machines has  decreased, whereas the number of particles per beam spill rises as is required for new treatments. In addition, details of specific designs are given in the main text and briefly summarised in Table \ref{tab:machine_params} (with further detail in the Appendix).

  \begin{table*}[!bth]
	\setlength\tabcolsep{3.5pt}
	\centering
	\caption{Summary of parameters of ion therapy machines. Some also offer proton therapy, detail not given here. Proposals that were never constructed are given in italics. For a full version with extended detail including citations, see Table \ref{tab:long_machine_params}.}
	\label{tab:machine_params}
	\begin{tabular}{llllclrll}
		\toprule
		\textbf{Name} & \textbf{Location} & \textbf{Active}  & \textbf{Extraction} & \textbf{Circ. (m)}  & \textbf{Main Tech.} & \textbf{Species} & \textbf{Extracted}   & \textbf{Particles} \\
		              &                   & \textbf{Years}   & \textbf{Method}     &                     &                  &                  & \textbf{KE (MeV/u)} & \textbf{per Spill} \\
		\midrule
		HIMAC         & Japan             & 1994 -           & Resonant            & \num{130}(*\num{2}  &  Synchrotron     & He               & \num{100}-\num{800}     & \num{1.2E10}             \\
		              &                   &                  &                     & rings)              &                  & C                & \num{100}-\num{800}     & \num{2.0E9}              \\
		GSI           & Germany           & 1998 - 2008      & Resonant            & \num{216.7}         &  Synchrotron     & C                & \num{80}-\num{430}      & \num{1.0E8}              \\
		HIBMC         & Japan             & 2003 -           & Resonant            & \num{94}            &  Synchrotron     & C                & \num{70}-\num{320}      & \num{2.0E9}              \\
\textit{iRCMS}        & Proposal          & ---              & ---                 & \num{60}            &  Synchrotron     & C                & \num{96}-\num{450}      & \num{2.7E7}             \\
        HIT           & Germany           & 2009 -           & RF-KO               & \num{65}            &  Synchrotron     & He               & \num{51}-\num{221}      & \num{1.0E10}             \\
                      &                   &                  &                     &                     &                  & C                & \num{88}-\num{430}      & \num{1.0E9}              \\
                      &                   &                  &                     &                     &                  & O                & \num{103}-\num{430}     & \num{5.0E8}              \\
        GHMC          & Japan             & 2009 -           & RF-KO               & \num{63}            &  Synchrotron     & C                & \num{140}-\num{400}     & \num{1.3E9}              \\
        HIRFL-CSR     & China             & 2009 -           & Fast: Kicker        & \num{161}           &  Synchrotron     & C                & \num{100}-\num{430}     & \num{5.0E8}              \\
                      &                   &                  & Slow: Resonant      &                     &                  &                  &                         &                    \\
\textit{PAMELA}       & Proposal          & ---              & Kicker              & \num{58}            &  FFA             & C                & \num{110}-\num{440}     & \num{5.0E6}              \\
        CNAO          & Italy             & 2010 -           & RF-KO               & \num{77.6}          &  Synchrotron     & C                & \num{120}-\num{400}     & \num{1.5E9}              \\
        MIT           & Germany           & 2015 -           & RF-KO               & \num{65}            &  Synchrotron     & C                & \num{85}-\num{430}      & \num{7.0E8}              \\
\textit{ARCHADE}      & Proposal          & ---              & Deflector           & \num{21}            &  Cyclotron       & C                & \num{400}               & ---                \\
\textit{KHIMA}        & Proposal          & ---              & Resonant            & \num{75}            &  Synchrotron     & C                & \num{110}-\num{430}     & \num{1.4E9}              \\
        HIMM Wuwei    & China             & 2019 -           & RF-KO               & \num{56.2}          &  Synchrotron     & C                & \num{120}-\num{400}     & \num{4.0E9}              \\
\textit{Quantum}      & Proposal          & ---              & ---                 & \num{28}            &  Laser and       & C                & \num{56}-\num{430}      & \num{1.0E8}             \\
\textit{Scalpel}      &                   &                  &                     &                     &  Synchrotron     &                  &                         &                   \\
\textit{NIMMS}        & Proposal          & ---              & Fast: Kicker        & NC: \num{76}        &  Synchrotron     & He               & \num{60}-\num{250}      & \num{8.2E10}            \\
                      &                   &                  & Slow: RF-KO         & DBA: \num{55}       &                  & C                & \num{100}-\num{430}     & \num{2.0E10}            \\
                      &                   &                  &                     & SC: \num{27}        &                  & O                & \num{100}-\num{430}     & \num{1.4E10}            \\
		\bottomrule
	\end{tabular}
\end{table*}

\subsection{Synchrotrons}
The design choice for most facilities is based on a synchrotron over a cyclotron, as it is capable of acceleration of particles with higher magnetic rigidity than protons, and the beam can be extracted over a wide range of energies, avoiding losses due to energy modulation \cite{Gerbershagen2016}.

The first hadron therapy treatments were delivered at the Bevelac \cite{Castro1993}. Hadron therapy was also delivered at GSI from \num{1997} to \num{2007}; treatment there was taken over by HIT \cite{Schardt}.

The first accelerator purpose-built for heavy ion therapy was the HIMAC at NIRS in Japan, which began treatment in \si{1994} \cite{Yamada1996}. The synchrotron has two separate, identical rings, each able to deliver ions from \num{100}$-$\SI{800}{\mega\electronvolt/u}. Several similar synchrotrons were then developed to achieve the same clinical requirements with a smaller machine and lower costs. One option was the PATRO project, resulting in the HIBMC where carbon ion treatment started in \num{2003} \cite{Murakami2009}. 
Both machines deliver to smaller energy ranges than HIMAC, reducing the circumference. A more compact alternative at NIRS led to GHMC ($^{12}$C only), which used the same focusing structure but increased the number of dipoles to three per cell \cite{Noda2007}. The GHMC design has been the basis for the most recent generation of Japanese ion therapy centres.

In Europe, HIT has offered full 3D raster scanning since treatments began in 2009, delivering protons, carbon and heavier ions \cite{Ondreka2008}. This was supplemented by the Proton-Ion Medical Machine Study (PIMMS), a collaboration to provide a strong baseline for future cancer therapy synchrotrons \cite{Bryant1999}. The lattice uses triplet focusing, where the dispersion-free straights are used for injection, extraction, and acceleration by RF; the extraction method uses a betatron core. Two machines were subsequently built from the PIMMS design at CNAO (Italy) and MedAustron (Austria).

The successor to PIMMS, the Next Ion Medical Machine Study (NIMMS), is currently ongoing, alongside the SEEIIST ion therapy centre proposal \cite{Damjanovic2021, Benedetto2021}. The aim is to offer a range of ions with an order of magnitude more particles per spill, while also being smaller than its predecessors. It is expected to offer slow extraction with multiple energies per spill, and fast extraction at microsecond timescales. This requires extensive R\&D, with three main proposals: a compact synchrotron using normal-conducting (NC) magnets; an even smaller synchrotron using superconducting (SC) magnets; and a `full linac' design. Of these, the linac option requires the most R\&D \cite{Benedetto2021}. 
The NC synchrotron has two options: 1) a modification of the PIMMS design with improved injection and extraction, or 2) a novel double bend achromat (DBA) design with a smaller circumference \cite{Zhang2021}. The SC proposal \cite{Benedetto2021arXiv} follows ideas evolved at NIRS \cite{Iwata2018}, using large-angle magnets with canted cosine theta (CCT) geometry \cite{Goodzeit2003} and nested alternating gradient (AG) coils \cite{Wan2015}. Although this option is more compact, development is likely to take considerable time.

There has also been some interest in an ion Rapid Cycling Medical Synchrotron (iRCMS), which allows for more rapid energy variation assuming single energy extraction \cite{Satogata2006}. Other than the higher magnet ramp rate, the iRCMS would have similar parameters to other synchrotron options.

\subsection{Alternatives to Synchrotrons}
Though synchrotrons are the current workhorse for heavy ion therapy, there are alternatives that could be implemented. For example, a therapeutic heavy ion cyclotron has been proposed at ARCHADE in France \cite{Revol2010, Balosso2020}, potentially beginning treatments in \num{2023}. Developments in high gradient cavities have made linacs more promising. As well as NIMMS, designs have also been presented by AVO-ADAM \cite{Degiovanni2016}(p$^{+}$) and ANL (p$^{+}$, $^{12}$C); the latter is still in its design phase \cite{Mustapha2019}. A `bent' full-linac for carbon ions has also been designed \cite{Bencini2020}. CABOTO, an NC `cyclinac', is a fast-cycling linac design, but has not yet been built due to the requirement of multiple high frequency klystron systems \cite{Andres}.

The time-independent magnets of Fixed Field Accelerators (FFAs) allow for a higher cycle rate, which could increase the number of particles delivered without requiring more particles per spill. However, FFAs tend to be larger than equivalent synchrotron counterparts, and magnet designs can become complicated. Multiple proposals for hadron therapy FFAs exist: the most-developed of these was PAMELA \cite{Peach2013}, which required two rings to accelerate protons and carbon ions over their full energy ranges (see Table~\ref{tab:machine_params}), but had a much higher cycle rate (up to \SI{1}{\kilo\hertz}) than equivalent synchrotrons (<\SI{1}{\hertz}). Although designs continue to advance, none have been constructed for treatment, and R\&D is required in areas such as beam stability and extraction.

A laser-hybrid accelerator known as `Quantum Scalpel' \cite{Shirai2018} is under development by Japanese industries working with QST-NIRS. This proposes a laser accelerator for low energies, and a SC synchrotron as the second acceleration stage. In the first stage, a Petawatt laser is incident on a thin target, producing an ion beam with low mean energy but broad energy spread \cite{Noda2019}. LhARA is a similar proposal \cite{Aymar2020}.

\subsection{Gantries}
The gantry is one of the key components of therapeutic beam delivery systems. It bends and focuses the beam in the plane perpendicular to a patient to deliver a precise dose to the treatment volume \cite{Owen2014, Linz2012}, with a typical momentum acceptance of ($\pm~0.5-1~\%$)$\Delta{p}/{p}$ \cite{Rizzoglio2020}. The majority of proton gantries use NC magnets. As carbon ions have three times the rigidity of protons, a suitable gantry's size and magnet weight increases dramatically; the first carbon ion gantry (at HIT) weighs \SI{600}{\tonne} and measures \SI{25}{\meter} \cite{Fuchs2017}, making installation and integration a challenge. Though it can transport fully-stripped carbon ions in the range \si{48}-\SI{430}{\mega\electronvolt/u} \cite{Galonska2013}, an alternative solution is required to shrink the gantry and increase momentum acceptance to deliver flexible, efficient treatment. One option is to use SC magnets: HIMAC employs six \SI{2.88}{\tesla} combined-function SC magnets, reducing gantry mass to $<$~\SI{300}{\tonne} and the length to \SI{13}{\meter} \cite{Iwata2018}; a developing design will incorporate \SI{5}{\tesla} magnets, reducing the gantry radius to $<$~\SI{5}{\meter} \cite{Iwata2018}. A recent design by TERA/CERN proposes a \SI{35}{\tonne} gantry\cite{Benedetto2021arXiv}, utilising \SI{90}{\degree}, \SI{4}{\tesla} AG-CCT dipoles \cite{Brouwer2019} to reduce the radius to \SI{5}{\meter}. These magnets are similar to those used for the NIMMS SC synchrotron.

Another path is to modify the gantry configuration. An example is the `GaToroid' (in its R\&D phase), which replaces the rotating beam transfer line with a large toroidal field, combined with a `vector' (steering) magnet to bend the beams from several directions towards the patient isocentre \cite{Bottura2020, Felcini2020}. One could also use high momentum acceptance FFA-style magnets \cite{Trbojevic2011}, though no such gantries have yet been constructed. A final solution may be a mounted gantry, similar to one constructed by Mevion (p$^+$) that rotates around the patient isocentre \cite{Zhao2016}. 

\section{Research Directions}

The main targets of hadron therapy R\&D are: cost and size reduction \cite{Gerbershagen2016}; improved reliability; new treatment capability.

\subsection{Lattice Design}
To reduce the size of accelerators and gantries, the options are either to change the structure and/or use SC technology. An example of the former is the NIMMS baseline lattice, which could be shrunk by using a DBA. Further reductions are limited by the necessity of including long drifts for extraction, while also providing enough space for the required bending elements. Conversely, the adoption of SC magnets uses similar beam optics, but the larger fields allow for tighter bending radii, significantly reducing the overall size.

\subsection{Magnet Design}
The use of SC magnets in accelerator and gantry design can increase the maximum field strength. Using combined-function magnets also reduces the total number of magnets. This can lead to cost reductions for the mechanical support structures and construction, but must be evaluated against the price of the cooling system and materials used for the magnets \cite{Gerbershagen2016}. 
To provide the most benefit, the system's magnetic composition (SC or hybrid NC/SC), configuration (e.g. CCT) and material (e.g. Nb-Ti) must be considered. CCT Nb-Ti magnets are being developed for future machines \cite{Benedetto2021}. 

\subsection{Extraction Methods}
Most ion therapy facilities now achieve slow extraction by excitation of a third-order resonance in combination with RF-knockout (RF-KO), as it can be used to extract multiple beam energies over the course of a single cycle \cite{Savazzi2019}. This allows different energy layers to be delivered rapidly, however the energy switching time is too long for treatments such as FLASH \cite{Jolly2020}. Some also use a kicker for fast extraction, delivering the entire stored beam in a single turn. The higher dose rates this provides may be useful for FLASH, but will be insufficient if rapid energy variation is also required.

One consideration for future extraction methods is the possibility to deliver multiple ions in a single treatment. This may be desirable to achieve a more conformal dose than with a single species \cite{Mairani2020}, or for novel imaging methods such as particle tomography and range verification \cite{Mazzucconi2018}. For multi-ion treatments, the issue is similar to that of rapid energy switching, and it is not yet clear what technologies will provide the required advances. A proposed SC gantry may be capable of performing proton tomography, so long as it has sufficiently large momentum acceptance \cite{Oponowicz2017}.

\subsection{Pre-Acceleration}
Almost all current ion therapy accelerators use a linac to accelerate ions before injection, although some facilities (in China) opt for a cyclotron \cite{Yuan2013}. Future pre-accelerators may need to transmit larger currents and reach higher energies to avoid space-charge effects in high intensity beams, such as those required for FLASH therapy \cite{Jolly2020}. 

The main objectives for new pre-accelerator systems are to reduce cost and size, while maintaining or increasing beam current. These depend on the ion source used; many facilities use the Supernanogan \cite{Pantechnik}, an ECR ion source, although higher current options such as AISHa \cite{Celona2019} or TwinEBIS \cite{Wenander2017} may come into greater use. An EBIS can produce small emittance beams, allowing for higher current through multi-turn injection, but cannot yet match the ECR in reliability and intensity. The linac itself will likely move to higher gradients, but the low duty cycle of injectors for medical synchrotrons makes SC options less favourable \cite{Benedetto2021}.

Novel ion production and pre-acceleration methods may eventually take over from linacs. In particular, laser-accelerator methods discussed earlier may provide beams of sufficient energy and intensity for clinical use. However, these technologies require further development; improvements on traditional methods will likely be used for the next generation of ion therapy machines.

\printbibliography

\section{Appendix}

In Table \ref{tab:long_machine_params} below, we give a full version of the ion therapy machine summary. This includes additional information on each accelerator, as well as some other facilities based on the same base machine design, and full citations for all the information.


\newpage
\begin{landscape}

  \begin{table*}[!htb]
	\setlength\tabcolsep{3.5pt}
	\captionsetup{oneside,margin={-7.0cm,0cm}}
	\caption{Parameters of ion therapy machines.  Entries that are proposals but never constructed are given in italics. As well as the sources given in the table, additional information is available from \cite{Kitagawa2010} and \cite{ptcog}.}
	\label{tab:long_machine_params}
	\hspace*{-8.35cm} 
	\begin{tabular}{llllllllllllrlll}
		\toprule
		\textbf{Name} & \textbf{Location} & \textbf{Active}  & \textbf{Main Tech.} & \textbf{Pre-Acc} & \textbf{Extraction} & \textbf{Rep. Rate} & \textbf{Circ. (m)} &\textbf{B$_\text{max}$ (T)}  & \textbf{RF Freq.} & \textbf{V$_\text{cavity}$}  & \textbf{Offer}   & \textbf{Species} & \textbf{Extracted}      & \textbf{Particles} & \textbf{Sources}                                               \\
		              &                   & \textbf{Years}   &                     &                  & \textbf{Method}     & \textbf{(Hz)}      &                    &                             & \textbf{Range}    & \textbf{Max (kV)}           & \textbf{Protons} &                  & \textbf{KE (MeV/u)}     & \textbf{per Spill} &                                                                \\
		\midrule                                                                    
		HIMAC         & Japan             & 1994 -           &  Synchrotron        & Linac            & Resonant            & 1.5                & \num{130}(*\num{2} & 1.5                         & 1.0-8.0           & 10                          & No               & He               & \num{100}-\num{800}     & \num{1.2E10}       &  \cite{Yamada1996}                                             \\
		              &                   &                  &                     &                  &                     &                    & rings)             &                             &                   &                             &                  & C                & \num{100}-\num{800}     & \num{2.0E9}        &                                                                \\
		GSI           & Germany           & 1998 - 2008      &  Synchrotron        & Linac            & Resonant            & 0.3                & \num{216.7}        & 1.8                         & 0.85-6.0          & 16                          & Yes              & C                & \num{80}-\num{430}      & \num{1.0E8}        &   \cite{Franczak1987, Schardt, Eickhoff}                       \\
		HIBMC         & Japan             & 2003 -           &  Synchrotron        & Linac            & Resonant            & 0.5                & \num{94}           & 1.38                        & 0.99-6.42         & 5.2                         & Yes              & C                & \num{70}-\num{320}      & \num{2.0E9}        &    \cite{Murakami2009, Hishikawa2002, Mishima2002, Itano}      \\
\textit{iRCMS}        & Proposal          & ---              &  Synchrotron        & Linac            & ---                 & 30                 & \num{60}           & 1.33                        & 0.65-3.57         & ---                         & No               & C                & \num{96}-\num{450}      & \num{2.7E7}        &   \cite{Trbojevic2011ircms, Satogata2006}                      \\
        HIT           & Germany           & 2009 -           &  Synchrotron        & Linac            & RF-KO               & 0.15               & \num{65}           & 1.53                        & 1.0-7.0           & 2                           & Yes              & He               & \num{51}-\num{221}      & \num{1.0E10}       &  \cite{Combs2010, Ondreka2008, Peters2016, Dolinskii2000}      \\
                      &                   &                  &                     &                  &                     &                    &                    &                             &                   &                             &                  & C                & \num{88}-\num{430}      & \num{1.0E9}        &                                                                \\
                      &                   &                  &                     &                  &                     &                    &                    &                             &                   &                             &                  & O                & \num{103}-\num{430}     & \num{5.0E8}        &                                                                \\
        GHMC$^1$      & Japan             & 2009 -           &  Synchrotron        & Linac            & RF-KO               & 0.36               & \num{63}           & 1.48                        & 0.9-7.0           & 2                           & No               & C                & \num{140}-\num{400}     & \num{1.3E9}        &   \cite{Ohno2011, Torikai2010, Noda2007}                       \\
        HIRFL-        & China             & 2009 -           &  Synchrotron        & Cyclotron        & Fast: Kicker        & 0.06               & \num{161}          & 1.4                         & 0.24-1.7          & 20                          & No               & C                & \num{100}-\num{430}     & \num{5.0E8}        &  \cite{Yuan2009, Yuan2013, Xia2002}                            \\
        CSR           &                   &                  &                     &                  & Slow: Resonant      &                    &                    &                             &                   &                             &                  &                  &                         &                    &                                                                \\
\textit{PAMELA}       & Proposal          & ---              &  FFA                & Cyclotron        & Kicker              & 1000               & \num{58}           & 4.25                        & 19.2-45.6         & 15                          & Yes              & C                & \num{110}-\num{440}     & \num{5.0E6}        &     \cite{Peach2013}                                           \\
        CNAO$^2$      & Italy             & 2010 -           &  Synchrotron        & Linac            & RF-KO               & 0.4                & \num{77.6}         & 1.5                         & 0.47-2.76         & 5                           & Yes              & C                & \num{120}-\num{400}     & \num{1.5E9}        &    \cite{Rossi2015, Rossi2006, Angoletta2008, Bryant2000}      \\
        MIT$^3$       & Germany           & 2015 -           &  Synchrotron        & Linac            & RF-KO               & 1                  & \num{65}           & 1.43                        & 1.0-7.0           & 2.5                         & Yes              & C                & \num{85}-\num{430}      & \num{7.0E8}        &   \cite{Rohdjess, Scheeler2016, Moller, moller2007synchrotron} \\
\textit{ARCHADE}      & Proposal          & ---              &  Cyclotron          & ---              & Deflector           & ---                & \num{21}           & 4.5                         & 75                & 80                          & No               & C                & \num{400}               & ---                &   \cite{Jongen2010, Balosso2020}                               \\
\textit{KHIMA}        & Proposal          & ---              &  Synchrotron        & Linac            & Resonant            & 0.3                & \num{75}           & 1.5                         & 0.49-5.83         & 0.79                        & Yes              & C                & \num{110}-\num{430}     & \num{1.4E9}        &    \cite{Yim2015, Kim2016}                                     \\
        HIMM$^4$      & China             & 2019 -           &  Synchrotron        & Cyclotron        & RF-KO               & 0.31               & \num{56.2}         & 1.66                        & 0.6-3.9           & 5                           & No               & C                & \num{120}-\num{400}     & \num{4.0E9}        &  \cite{Shi2019, Yang2014, zhang2016main}                       \\
\textit{Quantum}      & Proposal          & ---              &  Laser and          & Laser            & ---                 & ---                & \num{28}           & 4.0                         & 10                & ---                         & No               & C                & \num{56}-\num{430}      & \num{1.0E8}        &   \cite{Noda2019, Shirai2018, noda2019review}                  \\
\textit{Scalpel}      &                   &                  &  Synchrotron        &                  &                     &                    &                    &                             &                   &                             &                  &                  &                         &                    &                                                                \\
\textit{NIMMS}        & Proposal          & ---              &  Synchrotron        & Linac            & Fast: Kicker        & ---                & NC: \num{76}       & 1.5                         & ---               & ---                         & Yes              & He               & \num{60}-\num{250}      & \num{8.2E10}       &   \cite{Benedetto2021, Benedetto2021arXiv, Zhang2021}          \\
                      &                   &                  &                     &                  & Slow: RF-KO         &                    & DBA: \num{55}      & 1.5                         &                   &                             &                  & C                & \num{100}-\num{430}     & \num{2.0E10}       &                                                                \\
                      &                   &                  &                     &                  &                     &                    & SC: \num{27}       & 4.5                         &                   &                             &                  & O                & \num{100}-\num{430}     & \num{1.4E10}       &                                                                \\
		\bottomrule 
	\multicolumn{16}{l}{$^1$ Many facilities follow the GHMC design, mainly in Japan \cite{Iwata2018}. These include SAGA-HIMAT \cite{Kudo2013}, i-ROCK \cite{Nakayama2015} and Yamagata University Hospital} \\
	\multicolumn{16}{l}{$^2$ CNAO and MedAustron \cite{Pivi2019} share near identical parameters. Though it once used a betatron core for extraction, CNAO now uses RF-KO} \\
	\multicolumn{16}{l}{$^3$ MIT and SPHIC \cite{Jiang2014} are essentially the same design \cite{moller2007synchrotron}}  \\
	\multicolumn{16}{l}{$4$ The same design is being followed for both HIMM Wuwei and HIMM Lanzhou}
	\end{tabular}
\end{table*}
\end{landscape}


\end{document}